\documentclass[a4paper,11pt]{amsart}
\usepackage[utf8]{inputenc}
\usepackage{amsaddr}
\usepackage{calrsfs}
\usepackage{graphicx,color}
\usepackage[english]{babel}
\usepackage{graphicx}
\usepackage{amsmath}
\usepackage{epstopdf}
\usepackage{float}
\usepackage{color}

\usepackage{amssymb}

\numberwithin{equation}{section}

\theoremstyle{definition}

\theoremstyle{plain}
\newtheorem{thm}{Theorem}[section]

\newtheorem{cor}{Corollary}[section]

\theoremstyle{definition}

\newcommand{\E}{\mathbb{E}}

\begin{document}
\title[Fractional short rate ]
{Pricing European option with the short rate under Subdiffusive fractional Brownian motion regime}

\date{\today}

\author[Shokrollahi]{Foad Shokrollahi}
\address{Department of Mathematics and Statistics, University of Vaasa, P.O. Box 700, FIN-65101 Vaasa, FINLAND}
\email{foad.shokrollahi@uva.fi}

\begin{abstract}
The purpose of this paper is to analyze the problem of option pricing when the short rate follows subdiffusive fractional Merton model. We incorporate the stochastic nature of the short rate in our option valuation model and derive explicit formula for call and put option and discuss the corresponding fractional Black-Scholes equation. We present some properties of this pricing model for the cases of $\alpha$ and $H$. Moreover, the numerical simulations illustrate that our model is flexible and easy to implement.
\end{abstract}


\maketitle
\section{Introduction}\label{sec:0}

Nowadays, the Black–Scholes $(BS)$ model \cite{black1973pricing} is still classical and most popular model of the market. However, empirical research shows that
it cannot capture many of the characteristic features of prices, such as: long-range correlations, heavy-tailed and skewed
marginal distributions, lack of scale invariance, periods of constant values, etc. Therefore, improvements of the $BS$ model
itself did not stand still either. Since fractional Brownian motion $(FBM)$ has two important
properties called self-similarity and long-range dependence, it has the ability to capture the typical tail behavior of stock prices or indexes \cite{sottinen2003arbitrage,rogers1997arbitrage,cheridito2003arbitrage,necula2002option}. The $FBM$ model is an improvement of the $BS$ model, by replacing the $FBM$ with the Brownian motion in the
standard $BS$ model. That is
\begin{eqnarray}
\frac{dS_t}{S_t}=\mu dt+\sigma dB^H(t),
\label{eq:1}
\end{eqnarray}

here $\mu, \sigma$ are constants, and $B^H$ is a $FBM$ with Hurst parameter $H\in[\frac{1}{2},1)$.

Subdiffusive Brownian motion is an another generalization of the $BS$ model, which is introduced by Magdziarz \cite{magdziarz2009black}. In order to describe properly financial data exhibiting periods of constant
values, he put forward the subdiffusive strategy based on the geometric Brownian motion
to describe financial data with the periods of the constant prices. He
replaced the physical time $t$ with inverse $\alpha$-stable subordinator $T_{\alpha}(t)$ in the standard $BS$ model
where $\alpha\in (0,1)$. Magdziarz showed that the considered model is arbitrage-free but incomplete, and obtained the corresponding
subdiffusive $BS$ formula for the fair prices of European options. Moreover, Hui Gua et al. \cite{gu2012time} applied subdiffusive $FBM$ regime
\begin{eqnarray}
X_{\alpha}(t)=X(T_{\alpha}(t)),
\label{eq:2}
\end{eqnarray}

as the model of asset prices exhibiting subdiffusive dynamics. Here the parent process $X(\tau )$ is the $FBM$ defined in Equation (\ref{eq:1}),
$T_{\alpha}(t)$ is the inverse $\alpha$-stable subordinator with $\alpha\in (0,1)$. Later, many scholars made some improvements of this model \cite{gu2012time,wang2012continuous,hahn2011fokker}.

Constant short rate during the life of the option is the assumption at all above studies. This assumption is clearly at odds with reality because, as a matter of fact, the
short rate $r(t)$ is evolving random of time. Hence, in this study, we combine the stochastic nature
into our option pricing model. Specifically, we will consider the option pricing of the European options under the Merton short rate model \cite{merton1970dynamic} in a subdiffusive $FBM$ regime. That is, $r(t)=X(T_{\alpha}(t))$ in which $X(\tau )$ follows

\begin{eqnarray}
dX(\tau)=\mu_rd\tau+\sigma_r dB_1^{H}(\tau),
\label{eq:3}
\end{eqnarray}

and the stock price $S(t)=\widehat{X}(T_{\alpha}(t))$ in which $\widehat{X}(\tau )$ follows
\begin{eqnarray}
d\widehat{X}(\tau)=\mu_s\widehat{X}(\tau)d\tau+\sigma_s \widehat{X}(\tau)dB_2^{H}(\tau),
\label{eq:4}
\end{eqnarray}

where $\mu_r, \sigma_r, \mu_s, \sigma_s, $ are constant, $B_1^{H}(\tau)$ and $B_2^{H}(\tau)$ are two $FBM$ with Hurst parameter $H \in[\frac{1}{2}, 1)$ and correlation coefficient $\rho$. $T_{\alpha}(t)$ is the inverse $\alpha$-stable subordinator with $\alpha\in (0,1)$ defined as follows

\begin{eqnarray}
T_{\alpha}(t)=\inf\{\tau>0: U_{\alpha}(\tau)>t\},
\label{eq:5}
\end{eqnarray}

$\{U_{\alpha}(\tau)\}_{\tau\geq0}$ is a $\alpha$-stable Levy process with nonnegative increments and Laplace transform: $E\left(e^{-uU_{\alpha}(\tau)}\right)=e^{-\tau u^{\alpha}}$.

Fig. \ref{fig1} shows typically the differences and relationships between the sample paths of the stock price in the $FBM$ model and the subdiffusive $FBM$ model.

\begin{figure}[H]
  \centering
          \includegraphics[width=1\textwidth]{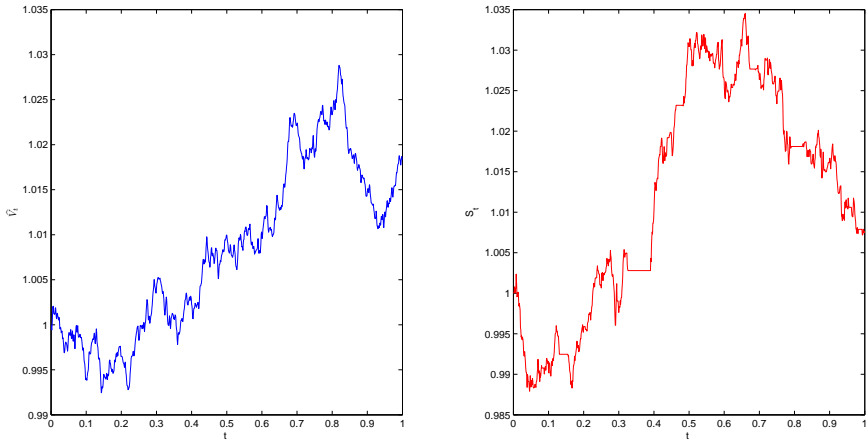}
  \caption{Comparison of the sample paths of the stock price in the $FBM$ model (left) and the subdiffusive $FBM$ model (right) for  $r=0.01, \alpha=0.9, H=0.8, \sigma=0.1, S_0=1$.}
\label{fig1}
\end{figure}

Here, assume that $T_{\alpha}(t)$ is independent of  $B_1^{H}(\tau)$ and $ B_2^{H}(\tau)$. Specially, when $H = \frac{1}{2}$, it is a subdiffusion process mentioned in Refs. \cite{magdziarz2009stochastic,magdziarz2010path} and when $\alpha\uparrow1$, $T_{\alpha}(t)$ reduces to physical time $t$. In this study, we apply the subdiffusive mechanism of trapping events in order to describe financial data exhibiting periods of constant values.

This paper is organized as follows. In Section \ref{sec:1}, we derive the formula for the price of a riskless zero-coupon
bond paying \$1 at maturity. In Section \ref{sec:2}, we obtain the corresponding $BS$ equation by using delta hedging argument and discuss some special cases of this equation. In Section \ref{sec:3}, we present an
analytic pricing formula for the European call and put options. In Section \ref{sec:4}, we study some special properties of this pricing formula. Furthermore, we show how to use our model to price options by numerical simulations. The comparison of our model and traditional models is undertaken in
this section. Finally, Section \ref{sec:5} draws the concluding remarks.

\section{Pricing formula for zero-coupon bond}\label{sec:1}
The purpose of this section is to derive the pricing formula for zero-coupon bond $P(r, t, T)$. Here, $P(r, T; T) = 1$, that is, the zero-coupon bond will pay
for 1 dollar at the expiry date $T$.

We assume that the short rate $r(t)$ satisfy Equation (\ref{eq:3}), $\alpha\in (\frac{1}{2}, 1)$ and $2\alpha-\alpha H>1$, then by applying the Taylor series expansion to $P(r, t, T)$ we obtain that

\begin{eqnarray}
P(r+\Delta r, t+\Delta t)&=& P(r,t,T)+\frac{\partial P}{\partial r}\Delta r+\frac{\partial P}{\partial t}\Delta t\nonumber\\
&&+\frac{1}{2}\frac{\partial^2 P}{\partial r^2}(\Delta r)^2++\frac{1}{2}\frac{\partial^2 P}{\partial r\partial t}\Delta r(\Delta t)+\frac{1}{2}\frac{\partial^2 P}{\partial t^2}(\Delta t)^2+O(\Delta t).
\label{eq:6}
\end{eqnarray}
From, Equation (\ref{eq:3}) and \cite{wang2012continuous}, we have

\begin{eqnarray}
\Delta r&=&\mu_r(\Delta T_{\alpha}(t))+\sigma_rB_1^H(T_{\alpha}(t))\nonumber\\
&=&\mu_r\left(\frac{t^{\alpha-1}}{\Gamma(\alpha)}\right)^{2H}(\Delta t)^{2H}+\sigma_r\Delta B_1^H(T_{\alpha}(t))+O((\Delta t)^{2H}).\\
(\Delta r)^2&=&\sigma_r^2\left(\frac{t^{\alpha-1}}{\Gamma(\alpha)}\right)^{2H}(\Delta t)^{2H}+O((\Delta t)^{2H}).\\
\Delta r(\Delta t)&=&O((\Delta t)^{2H}).
\label{eq:7}
\end{eqnarray}

Then from the Lemma 1 in \cite{wang2012continuous}, we can get

\begin{eqnarray}
dP(r, t ,T)&=&\left[\left(\frac{t^{\alpha-1}}{\Gamma(\alpha)}\right)^{2H}\left(\mu_r\frac{\partial P}{\partial r}+\frac{1}{2}\sigma_r^2\frac{\partial^2 P}{\partial r^2}\right)2Ht^{2H-1}+\frac{\partial P}{\partial t}\right]dt\nonumber\\
&&+\sigma_r\frac{\partial P}{\partial t}dB_1^H(T_{\alpha}(t)).
\label{eq:8}
\end{eqnarray}
Assuming

\begin{eqnarray}
\mu&=&\frac{1}{P}\left[\left(\frac{t^{\alpha-1}}{\Gamma(\alpha)}\right)^{2H}\left(\mu_r\frac{\partial P}{\partial r}+\frac{1}{2}\sigma_r^2\frac{\partial^2 P}{\partial r^2}\right)2Ht^{2H-1}+\frac{\partial P}{\partial t}\right],\nonumber\\
\sigma&=&\frac{1}{P}\left(\frac{\partial P}{\partial r}\right),
\label{eq:9}
\end{eqnarray}
and letting the local expectations hypothesis holds for the term structure of interest rates
(i.e. $\mu=r$), we obtain
\begin{eqnarray}
&&\frac{\partial P}{\partial t}+2Ht^{2H-1}\mu_r\left(\frac{t^{\alpha-1}}{\Gamma(\alpha)}\right)^{2H}\frac{\partial P}{\partial r}\nonumber\\
&&+Ht^{2H-1}\sigma_r^2\left(\frac{t^{\alpha-1}}{\Gamma(\alpha)}\right)^{2H}\frac{\partial^2 P}{\partial r^2}-rP=0.
\label{eq:10}
\end{eqnarray}
Then, zero-coupon bond $P(r, t, T)$ with boundary condition $P(r, t, T)=1$ satisfy the following partial differential equation
\begin{eqnarray}
&&\frac{\partial P}{\partial t}+2Ht^{2H-1}\mu_r\left(\frac{t^{\alpha-1}}{\Gamma(\alpha)}\right)^{2H}\frac{\partial P}{\partial r}\nonumber\\
&&+Ht^{2H-1}\sigma_r^2\left(\frac{t^{\alpha-1}}{\Gamma(\alpha)}\right)^{2H}\frac{\partial^2 P}{\partial r^2}-rP=0.
\label{eq:11}
\end{eqnarray}

To solve Equation (\ref{eq:11}) for $P(r, t, T)$, let $\tau =T-t, P(r, t, T) = \exp\{f_1(\tau)-rf_2(\tau)\} $, then  we have
\begin{eqnarray}
\frac{\partial P}{\partial t}&=&P\left(-\frac{\partial f_1(\tau)}{\partial t}+r\frac{\partial f_2(\tau)}{\partial t}\right),
\label{eq:12}
\end{eqnarray}
\begin{eqnarray}
\frac{\partial P}{\partial r}&=&-Pf_2(\tau),
\label{eq:13}
\end{eqnarray}
\begin{eqnarray}
\frac{\partial^2 P}{\partial r^2}&=&Pf_2(\tau)^2.
\label{eq:14}
\end{eqnarray}
Substituting Equations (\ref{eq:13}) and (\ref{eq:14}) into Equation (\ref{eq:12}) and simplifying Equation (\ref{eq:11}) becomes
\begin{eqnarray}
&&P\Bigg[Ht^{2H-1}\sigma_r^2f_2(\tau)^2\left(\frac{t^{\alpha-1}}{\Gamma(\alpha)}\right)^{2H}-2Ht^{2H-1}\mu_rf_2(\tau)\left(\frac{t^{\alpha-1}}{\Gamma(\alpha)}\right)^{2H}\nonumber\\&&-\frac{\partial f_1(\tau)}{\partial \tau}
+r\left(\frac{\partial f_2(\tau)}{\partial t}-1\right)\Bigg]=0.
\label{eq:15}
\end{eqnarray}
From Equation (\ref{eq:15}), we have
\begin{eqnarray}
\frac{\partial f_1(\tau)}{\partial \tau}&=&Ht^{2H-1}\left(\frac{t^{\alpha-1}}{\Gamma(\alpha)}\right)^{2H}\left(\sigma_r^2f_2(\tau)^2-2\mu_rf_2(\tau)\right),\nonumber\\
\frac{\partial f_2(\tau)}{\partial \tau}&=&1.
\label{eq:16}
\end{eqnarray}

Then,

\begin{eqnarray}
f_1(\tau)&=&\frac{H\sigma_r^2}{(\Gamma(\alpha))^{2H}}\int_0^{\tau}(T-s)^{(\alpha-1)2H+2H-1}s^2ds\nonumber\\
&&-\frac{2H\mu_r}{(\Gamma(\alpha))^{2H}}\int_0^{\tau}(T-s)^{(\alpha-1)2H+2H-1}sds,\label{eq:16-1}\\
f_2(\tau)&=&\tau.
\label{eq:17}
\end{eqnarray}

Hence, we obtain a formula for the price at time $t$ of a riskless zero-coupon bond which
pay \$1 at maturity $T$ is given by
\begin{eqnarray}
P(r, t, T)=e^{-r\tau+f_1(\tau)}.
\label{eq:18}
\end{eqnarray}

\begin{cor}

When $\alpha\uparrow1$, Equations (\ref{eq:3}) and (\ref{eq:4}) reduce to the $FBM$, we obtain
\begin{eqnarray}
f_1(\tau)&=&H\sigma_r^2\int_0^{\tau}(T-s)^{2H-1}s^2ds-2H\mu_r\int_0^{\tau}(T-s)^{2H-1}sds,
\label{eq:18-1}
\end{eqnarray}
specially, if $t=0$
\begin{eqnarray}
f_1(\tau)&=&\sigma_r^2\frac{T^{2H+2}}{(2H+1)(2H+2)}-\mu_r\frac{T^{2H+1}}{2H+1},
\label{eq:18-2}
\end{eqnarray}
then

\begin{eqnarray}
P(r, t, T)=\exp\left\{-rT+\sigma_r^2\frac{T^{2H+2}}{(2H+1)(2H+2)}-\mu_r\frac{T^{2H+1}}{2H+1}\right\}.
\label{eq:18-3}
\end{eqnarray}
\label{cor:1}
\end{cor}

\begin{cor}
If $H=\frac{1}{2}$, from Equation (\ref{eq:16-1}), we obtain

\begin{eqnarray}
f_1(\tau)&=&\frac{1}{2}\frac{\sigma_r^2}{\Gamma(\alpha)}\int_0^{\tau}(T-s)^{\alpha-1}s^2ds\nonumber\\
&&-\frac{\mu_r}{\Gamma(\alpha)}\int_0^{\tau}(T-s)^{\alpha-1}sds,
\label{eq:19}
\end{eqnarray}

then the result is consistent with the result in \cite{guo2017option}.

Further, if $\alpha\uparrow1$ and $H=\frac{1}{2}$, Equations (\ref{eq:3}) and (\ref{eq:4}) reduce to the geometric Brownian motion, then we have

\begin{eqnarray}
f_1(\tau)=\frac{1}{6}\sigma_r^2\tau^3-\frac{1}{2}\mu_r\tau^2,
\label{eq:20}
\end{eqnarray}
then
\begin{eqnarray}
P(r, t, T)=e^{-r\tau+\frac{1}{6}\sigma_r^2\tau^3-\frac{1}{2}\mu_r\tau^2}.
\label{eq:21}
\end{eqnarray}
which is consistent with the result in \cite{kung2009option,cui2010comment}.
\label{cor:2}
\end{cor}
\section{Fractional $BS$ equation}\label{sec:2}

The purpose of this section is to derive the fractional $BS$ equation for European options when the short rate $r(t)$ and stock price  $S(T)=S(T_{\alpha}(t)$ satisfy Equations (\ref{eq:3}) and (\ref{eq:4}), respectively. We assume that $B_1^{H}(T_{\alpha}(t))$ and $B_2^{H}(T_{\alpha}(t))$ are two $FBM$ with Hurst parameter $H \in[\frac{1}{2}, 1)$ and correlation coefficient $\rho$.

Let $C=C(S, r,t)$ be the price of a European call option at time t with a strike price $K$ that matures at time $T$. Then we
have.

\begin{thm}
Assume that the stock price short rate $r(t)$ and $S(t)$ satisfy Equations (\ref{eq:3}) and (\ref{eq:4}), respectively. Then, $C(S, r,t)$  satisfies the following fractional $BS$ equation
\begin{eqnarray}
&&\frac{\partial C}{\partial t}+\widetilde{\sigma}_s^2(t)S^2\frac{\partial^2 C}{\partial S^2}+\widetilde{\sigma}_r^2(t)\frac{\partial^2 C}{\partial r^2}+2\rho\widetilde{\sigma}_r(t)\widetilde{\sigma}_s(t)\frac{\partial^2 C}{\partial S\partial r}\nonumber\\
&&+2Ht^{2H-1}\mu_r\left(\frac{t^{\alpha-1}}{\Gamma(\alpha)}\right)^{2H}\frac{\partial C}{\partial r}+rS\frac{\partial C}{\partial S}-rC=0,
\label{eq:22}
\end{eqnarray}
where
\begin{eqnarray}
\widetilde{\sigma}_s^2(t)=Ht^{2H-1}\sigma_s^2\left(\frac{t^{\alpha-1}}{\Gamma(\alpha)}\right)^{2H},
\label{eq:23}
\end{eqnarray}
\begin{eqnarray}
\widetilde{\sigma}_r^2(t)=Ht^{2H-1}\sigma_r^2\left(\frac{t^{\alpha-1}}{\Gamma(\alpha)}\right)^{2H}.
\label{eq:24}
\end{eqnarray}
$\sigma_s, \sigma_r, \mu_s, \mu_s,$ are constant, $H\in[\frac{1}{2}, 1)$ and $\alpha\in (\frac{1}{2}, 1)$ and $2\alpha-\alpha H>1$.
\label{th:1}
\end{thm}
\textbf{Proof:}
We consider a portfolio with $D_{1t}$ units of stock and $D_{2t}$ units of zero-coupon bond $P(r, t, T)$ and one unit of $C=C(r, t, T)$. Then,
the value of the portfolio at current time $t$ is
\begin{eqnarray}
\Pi_t=C-D_{1t}S_t-D_{2t}P_t.
\label{eq:25}
\end{eqnarray}
Then, from \cite{guo2017option} we have
\begin{eqnarray}
d\Pi_t&=&C_t-D_{1t}dS_t-D_{2t}dP_t\nonumber\\
&=&\Bigg[\frac{\partial C}{\partial t}dt+Ht^{2H-1}\sigma_s^2S_t^2\left(\frac{t^{\alpha-1}}{\Gamma(\alpha)}\right)^{2H}\frac{\partial^2 C}{\partial S^2}+Ht^{2H-1}\sigma_r^2\left(\frac{t^{\alpha-1}}{\Gamma(\alpha)}\right)^{2H}\frac{\partial^2 C}{\partial r^2}\nonumber\\
&+&2Ht^{2H-1}\rho\sigma_r\sigma_sS\left(\frac{t^{\alpha-1}}{\Gamma(\alpha)}\right)^{2H}\frac{\partial^2 C}{\partial S\partial r}\Bigg]dt+\Bigg[\frac{\partial C}{\partial t}-D_{1t}\Bigg]dS_t\nonumber\\
&+&\Bigg[\frac{\partial C}{\partial r}-D_{2t}\frac{\partial P}{\partial r}\Bigg]dr+D_{2t}\Bigg[\frac{\partial P}{\partial t}+Ht^{2H-1}\sigma_r^2\left(\frac{t^{\alpha-1}}{\Gamma(\alpha)}\right)^{2H}\frac{\partial^2 P}{\partial r^2}\Bigg]dt.
\label{eq:26}
\end{eqnarray}
By setting $D_{1t}=\frac{\partial C}{\partial S},\,D_{2t}=\frac{\frac{\partial C}{\partial r}}{\frac{\partial P}{\partial r}},$ to eliminate the stochastic noise, then

\begin{eqnarray}
d\Pi_t&=&\nonumber\\
&=&\Bigg[\frac{\partial C}{\partial t}+Ht^{2H-1}\left(\frac{t^{\alpha-1}}{\Gamma(\alpha)}\right)^{2H}\left(\sigma_s^2S^2\frac{\partial^2 C}{\partial S^2}+\sigma_r^2\frac{\partial^2 C}{\partial r^2}+2\rho\sigma_r\sigma_sS\frac{\partial^2 C}{\partial S\partial r}\right)\Bigg]dt\nonumber\\
&-&\frac{\frac{\partial C}{\partial r}}{\frac{\partial P}{\partial r}}\Bigg[rP-2Ht^{2H-1}\mu_r\left(\frac{t^{\alpha-1}}{\Gamma(\alpha)}\right)^{2H}\frac{\partial P}{\partial r}\Bigg]dt.
\label{eq:27}
\end{eqnarray}

The return of an amount $\Pi_t$ invested in bank account
is equal to $r(t)\Pi_tdt$ at time $dt$, $\E(d\Pi_t)=r(t)\Pi_tdt=r(t)\left(C-D_{1t}S_t-D_{2t}P_t\right)$, hence from Equation (\ref{eq:27}) we have

\begin{eqnarray}
&&\frac{\partial C}{\partial t}+Ht^{2H-1}\left(\frac{t^{\alpha-1}}{\Gamma(\alpha)}\right)^{2H}\left(\sigma_s^2S^2\frac{\partial^2 C}{\partial S^2}+\sigma_r^2\frac{\partial^2 C}{\partial r^2}+2\rho\sigma_r\sigma_sS\frac{\partial^2 C}{\partial S\partial r}\right)\nonumber\\
&&+2Ht^{2H-1}\mu_r\left(\frac{t^{\alpha-1}}{\Gamma(\alpha)}\right)^{2H}\frac{\partial C}{\partial r}+rS\frac{\partial C}{\partial S}-rC=0.
\label{eq:28}
\end{eqnarray}
Let
\begin{eqnarray}
\widetilde{\sigma}_s^2(t)=Ht^{2H-1}\sigma_s^2\left(\frac{t^{\alpha-1}}{\Gamma(\alpha)}\right)^{2H},
\label{eq:29}
\end{eqnarray}
\begin{eqnarray}
\widetilde{\sigma}_r^2(t)=Ht^{2H-1}\sigma_r^2\left(\frac{t^{\alpha-1}}{\Gamma(\alpha)}\right)^{2H}.
\label{eq:30}
\end{eqnarray}
Then
\begin{eqnarray}
&&\frac{\partial C}{\partial t}+\widetilde{\sigma}_s^2(t)S_t^2\frac{\partial^2 C}{\partial S_t^2}+\widetilde{\sigma}_r^2(t)\frac{\partial^2 C}{\partial r^2}+2\rho\widetilde{\sigma}_r(t)\widetilde{\sigma}_s(t)\frac{\partial^2 C}{\partial S\partial r}\nonumber\\
&&+2Ht^{2H-1}\mu_r\left(\frac{t^{\alpha-1}}{\Gamma(\alpha)}\right)^{2H}\frac{\partial C}{\partial r}+rS\frac{\partial C}{\partial S}-rC=0,
\label{eq:31}
\end{eqnarray}

proof is completed.

From Theorem (\ref{th:1}), we can get the following corollaries

\begin{cor}

If $\rho=0$ and $r(t)$ be a constant, then the European call option $C=C(S, r, T)$ satisfies

\begin{eqnarray}
\frac{\partial C}{\partial t}+Ht^{2H-1}\sigma_s^2S_t^2\left(\frac{t^{\alpha-1}}{\Gamma(\alpha)}\right)^{2H}\frac{\partial^2 C}{\partial S_t^2}+rS\frac{\partial C}{\partial S}-rC=0,
\label{eq:32}
\end{eqnarray}
which is a fractional $BS$ equation considered in \cite{liang2012fractional}.

\end{cor}

\begin{cor}

When $\alpha\uparrow1$, we obtain

\begin{eqnarray}
&&\frac{\partial C}{\partial t}+Ht^{2H-1}\sigma_s^2S_t^2\frac{\partial^2 C}{\partial S_t^2}+Ht^{2H-1}\sigma_r^2\frac{\partial^2 C}{\partial r^2}+2 Ht^{2H-1}\rho\sigma_r\sigma_s\frac{\partial^2 C}{\partial S\partial r}\nonumber\\
&&+2Ht^{2H-1}\mu_r\frac{\partial C}{\partial r}+rS\frac{\partial C}{\partial S}-rC=0,
\label{eq:33}
\end{eqnarray}
Further, if $\rho=0$, $H=\frac{1}{2}$, and $r(t)$ be a constant, from Equation (\ref{eq:33}) we have the celebrated $BS$ equation
\begin{eqnarray}
\frac{\partial C}{\partial t}+\frac{1}{2}\sigma_s^2S_t^2\frac{\partial^2 C}{\partial S_t^2}+rS\frac{\partial C}{\partial S}-rC=0,
\label{eq:34}
\end{eqnarray}
\end{cor}

\section{Pricing formula under subdiffusive fractional Merton short rate model}\label{sec:3}
In this section, we propose an explicit formula for European call option when its value satisfy the partial differential equation (\ref{eq:22}) with boundary condition $C(S, r, T)=(S_T-K)^+$. Then, we can get
\begin{thm}
Let $r(t)$ satisfies Equation (\ref{eq:3}) and $S(t)$ satisfies Equation (\ref{eq:4}), then the price of  European call and put options with strike price $K$ and maturity $T$ are given by
\begin{eqnarray}
C(S, r, t)&=&S\phi(d_1)-KP(r, t, T)\phi(d_2),\\
P(S, r, t)&=&KP(r, t, T)\phi(-d_2)-\phi(-d_1).
\label{eq:35}
\end{eqnarray}
where
\begin{eqnarray}
d_1&=&\frac{\ln\frac{S}{K}-\ln P(r, t, T)+\frac{H}{(\Gamma(\alpha))^{2H}}\int_t^T\widehat{\sigma}^2(s)s^{(\alpha-1)2H+2H-1}ds}{\sqrt{\frac{2H}{(\Gamma(\alpha))^{2H}}\int_t^T\widehat{\sigma}^2(s)s^{(\alpha-1)2H+2H-1}ds}},
\label{eq:36}\\
d_2&=&d_1-\sqrt{\frac{2H}{(\Gamma(\alpha))^{2H}}\int_t^T\widehat{\sigma}^2(s)s^{(\alpha-1)2H+2H-1}ds},\\
\widehat{\sigma}^2(t)&=&\sigma_s^2+2\rho\sigma_r\sigma_s(T-t)+\sigma_r^2(T-t)^2.
\label{eq:37}
\end{eqnarray}
$P(r, t, T)$ is given by Equation (\ref{eq:18}) and $\phi(.)$ is the cumulative normal distribution function.
\label{th:2}
\end{thm}
\textbf{Proof:}

Consider the partial differential equation (\ref{eq:22}) of the European call option with boundary condition $C(S, r, T)=(S_T-K)^+$

\begin{eqnarray}
&&\frac{\partial C}{\partial t}+\widetilde{\sigma}_s^2(t)S_t^2\frac{\partial^2 C}{\partial S_t^2}+\widetilde{\sigma}_r^2(t)\frac{\partial^2 C}{\partial r^2}+2\rho\widetilde{\sigma}_r(t)\widetilde{\sigma}_s(t)\frac{\partial^2 C}{\partial S\partial r}\nonumber\\
&&+2Ht^{2H-1}\mu_r\left(\frac{t^{\alpha-1}}{\Gamma(\alpha)}\right)^{2H}\frac{\partial C}{\partial r}+rS\frac{\partial C}{\partial S}-rC=0.
\label{eq:38}
\end{eqnarray}
Denote
\begin{eqnarray}
z=\frac{S}{P(r, t, T)},\quad\Theta(z, t)=\frac{C(S, r, t)}{P(r, t, T)},
\label{eq:39}
\end{eqnarray}
therefore by computing, we get

\begin{eqnarray}
\frac{\partial C}{\partial t}&=&\Theta\frac{\partial P}{\partial t}+P\frac{\partial \Theta}{\partial t}-z\frac{\partial \Theta}{\partial z}\frac{\partial P}{\partial t},\nonumber\\
\frac{\partial C}{\partial r}&=&\Theta\frac{\partial P}{\partial r}-z\frac{\partial \Theta}{\partial z}\frac{\partial P}{\partial r},\nonumber\\
\frac{\partial C}{\partial S}&=&\frac{\partial \Theta}{\partial z},\label{eq:40}\\
\frac{\partial^2 C}{\partial r^2}&=&\Theta\frac{\partial^2 P}{\partial r^2}-z\frac{\partial \Theta}{\partial z}\frac{\partial^2 P}{\partial r^2}+\frac{z^2}{P}\frac{\partial^2\Theta}{\partial z^2}\left(\frac{\partial P}{\partial r}\right)^2,\nonumber\\
\frac{\partial^2 C}{\partial r\partial S}&=&-\frac{z}{P}\frac{\partial^2 \Theta}{\partial z^2}\frac{\partial P}{\partial r},\nonumber\\
\frac{\partial^2 C}{\partial S^2}&=&\frac{1}{P}\frac{\partial^2 \Theta}{\partial z^2}.\nonumber
\end{eqnarray}

Inserting Equation (\ref{eq:40}) into Equation (\ref{eq:38})

\begin{eqnarray}
\frac{\partial \Theta}{\partial t}&+&\frac{\partial^2 \Theta}{\partial z^2}\left[\widetilde{\sigma}_s^2(t)\frac{S^2}{P^2}+2\rho z^2 \widetilde{\sigma}_r(t)\widetilde{\sigma}_s(t)\frac{1}{P}\frac{\partial P}{\partial r}+\widetilde{\sigma}_r^2(t)z^2\left(\frac{1}{P}\frac{\partial P}{\partial r}\right)^2\right]\nonumber\\
&-&\frac{z}{P}\left[\frac{\partial P}{\partial t}+\widetilde{\sigma}_r^2(t)\frac{\partial^2 P}{\partial r^2}+2Ht^{2H-1}\mu_r\left(\frac{t^{\alpha-1}}{\Gamma(\alpha)}\right)^{2H}\frac{\partial P}{\partial r}-r\frac{S}{z}\right]\nonumber\\
&+&\frac{\Theta}{P}\left[\frac{\partial P}{\partial t}+\widetilde{\sigma}_r^2(t)\frac{\partial^2 P}{\partial r^2}+2Ht^{2H-1}\mu_r\left(\frac{t^{\alpha-1}}{\Gamma(\alpha)}\right)^{2H}\frac{\partial P}{\partial r}-rP\right]=0.
\label{eq:41}
\end{eqnarray}

From Equation (\ref{eq:11}), we can obtain

\begin{eqnarray}
\frac{\partial \Theta}{\partial t}+\overline{\sigma}^2(t)z^2\frac{\partial^2 \Theta}{\partial z^2}=0,
\label{eq:42}
\end{eqnarray}
with boundary condition $\Theta(z,T)=(z-K)^+$,

where
\begin{eqnarray}
\overline{\sigma}^2(t)=\widetilde{\sigma}_s^2(t)+2\rho\widetilde{\sigma}_r(t)\widetilde{\sigma}_s(t)(T-t)+\widetilde{\sigma}_r(t)^2(T-t)^2.
\label{eq:43}
\end{eqnarray}

The solution of partial differential Equation (\ref{eq:42}) with boundary condition $\Theta(z,T)=(z-K)^+$, is given by

\begin{eqnarray}
\Theta(z,t)=z\phi(\widehat{d}_1)-K\phi(\widehat{d}_2),
\label{eq:44}
\end{eqnarray}
here
\begin{eqnarray}
\widehat{d}_1&=&\frac{\ln\frac{z}{K}+\int_t^T\overline{\sigma}^2(s)ds}{\sqrt{2\int_t^T\widehat{\sigma}^2(s)ds}},\\
\widehat{d}_2&=&\widehat{d}_1-{\sqrt{2\int_t^T\overline{\sigma}^2(s)ds}}.
\label{eq:45}
\end{eqnarray}
Thus, from Equation (\ref{eq:39}) and (\ref{eq:44})-(\ref{eq:45}) we obtain

\begin{eqnarray}
C(S, r, t)=S\phi(d_1)-KP(r, t, T)\phi(d_2),
\label{eq:46}
\end{eqnarray}

where

\begin{eqnarray}
d_1&=&\frac{\ln\frac{S}{K}-\ln P(r, t, T)+\frac{H}{\left(\Gamma(\alpha)\right)^{2H}}\int_t^T\widehat{\sigma}^2(s)s^{(\alpha-1)2H+2H-1}ds}{\sqrt{\frac{2H}{\left(\Gamma(\alpha)\right)^{2H}}\int_t^T\widehat{\sigma}^2(s)s^{(\alpha-1)2H+2H-1}ds}},\\
d_2&=&d_1-\sqrt{\frac{2H}{\left(\Gamma(\alpha)\right)^{2H}}\int_t^T\widehat{\sigma}^2(s)s^{(\alpha-1)2H+2H-1}ds}.
\label{eq:47}
\end{eqnarray}

Letting $\alpha\uparrow1$, from Theorem \ref {th:2}, we obtain

\begin{cor}
Suppose that the short rate $r(t)$ satisfies Equation (\ref{eq:3}) and the stock price $S(t)$ satisfies Equation (\ref{eq:4}), then the price of  European call and put options with strike price $K$ and maturity $T$ are given by
\begin{eqnarray}
C(S, r, T)&=&S\phi(d_1)-KP(r, t, T)\phi(d_2),\\
P(S, r, T)&=&KP(r, t, T)\phi(-d_2)-S\phi(-d_1).
\label{eq:48}
\end{eqnarray}
where
\begin{eqnarray}
d_1&=&\frac{\ln\frac{S}{K}-\ln P(r, t, T)+H\int_t^T\widehat{\sigma}^2(s)s^{2H-1}ds}{\sqrt{2H\int_t^T\widehat{\sigma}^2(s)s^{2H-1}ds}},
\label{eq:48-1}\\
d_2&=&d_1-\sqrt{2H\int_t^T\widehat{\sigma}^2(s)s^{2H-1}ds},\\
\widehat{\sigma}^2(t)&=&\sigma_s^2+2\rho\sigma_r\sigma_s(T-t)+\sigma_r^2(T-t)^2,\\
P(r, t, T)&=&\exp\Bigg\{-r\tau+H\sigma_r^2\int_0^{\tau}(T-s)^{2H-1}s^2ds\nonumber\\
&&-2H\mu_r\int_0^{\tau}(T-s)^{2H-1}sds\Bigg\},\,\tau=T-t.
\label{eq:49}
\end{eqnarray}
More specifically, if $H=\frac{1}{2}$,  we have
\begin{eqnarray}
d_1&=&\frac{\ln\frac{S}{K}-\ln P(r, t, T)+\frac{1}{2}\varphi(t, T)}{\sqrt{\varphi(t, T)}},\\
d_2&=&d_1-\sqrt{\varphi(t, T)},\\
\varphi(t, T)&=&\sigma_s^2(T-t)+\rho \sigma_r\sigma_s(T-t)^2+\frac{1}{3}\sigma_r^2(T-t)^3,\\
P(r, t, T)&=&\exp\left\{-r(T-t)-\frac{1}{2}\mu_r(T-t)^2+\frac{1}{6}\sigma_r^2(T-t)^3\right\}.
\end{eqnarray}
which is consistent with result in \cite{cui2010comment}.
\end{cor}

Letting $H=\frac{1}{2}$, from Theorem \ref {th:2}, we can get

\begin{cor}
Suppose that the short rate $r(t)$ satisfies Equation (\ref{eq:3}) and the stock price $S(t)$ satisfies Equation (\ref{eq:4}), then the price of  European call and put options with strike price $K$ and maturity $T$ are given by
\begin{eqnarray}
C(S, r, T)&=&S\phi(d_1)-KP(r, t, T)\phi(d_2),\label{eq:49-1}\\
P(S, r, T)&=&KP(r, t, T)\phi(-d_2)-\phi(-d_1).
\label{eq:49-2}
\end{eqnarray}
where
\begin{eqnarray}
d_1&=&\frac{\ln\frac{S}{K}-\ln P(r, t, T)+\frac{1}{2\Gamma(\alpha)}\int_t^T\widehat{\sigma}^2(s)s^{\alpha-1}ds}{\sqrt{\frac{1}{\Gamma(\alpha)}\int_t^T\widehat{\sigma}^2(s)s^{\alpha-1}ds}},
\label{eq:49-3}\\
d_2&=&d_1-\sqrt{\frac{1}{\Gamma(\alpha)}\int_t^T\widehat{\sigma}^2(s)s^{\alpha-1}ds},\\
\widehat{\sigma}^2(t)&=&\sigma_s^2+2\rho\sigma_r\sigma_s(T-t)+\sigma_r^2(T-t)^2,\\
P(r, t, T)&=&\exp\Bigg\{-r\tau+\frac{\sigma_r^2}{2\Gamma(\alpha)}\int_0^{\tau}(T-s)^{\alpha-1}s^2ds\nonumber\\
&&-\frac{\mu_r}{(\Gamma(\alpha)}\int_0^{\tau}(T-s)^{\alpha-1}sds\Bigg\}.
\label{eq:50}
\end{eqnarray}

Specially, If $\rho=0$, from Equations (\ref{eq:49-1})-(\ref{eq:50}), we have
\begin{eqnarray}
d_1&=&\frac{\ln\frac{S}{K}-\ln P(r, t, T)+\frac{1}{2\Gamma(\alpha)}\int_t^T\widehat{\sigma}^2(s)s^{\alpha-1}ds}{\sqrt{\frac{1}{\Gamma(\alpha)}\int_t^T\widehat{\sigma}^2(s)s^{\alpha-1}ds}},
\label{eq:51}\\
d_2&=&d_1-\sqrt{\frac{1}{\Gamma(\alpha)}\int_t^T\widehat{\sigma}^2(s)s^{\alpha-1}ds},\\
\widehat{\sigma}^2(t)&=&\sigma_s^2+\sigma_r^2(T-t)^2,\\
P(r, t, T)&=&\exp\Bigg\{-r\tau+\frac{1}{2}\frac{\sigma_r^2}{\Gamma(\alpha)}\int_0^{\tau}(T-s)^{\alpha-1}s^2ds\nonumber\\
&&-\frac{\mu_r}{\Gamma(\alpha)}\int_0^{\tau}(T-s)^{\alpha-1}sds\Bigg\}.
\label{eq:52}
\end{eqnarray}
which is similar with results mentioned in \cite{guo2017option}.
\end{cor}

\section{Simulation studies}\label{sec:4}
Let us first discuss about the implied volatility of the subdiffusive $FBM$ model, then we will show some simulation findings.

\begin{cor}
If $t=0$, the value of European call option $\overline{C}(K, T)$ and put option $\overline{P}(K, T)$ can be written as
\begin{eqnarray}
\overline{C}(K, T)&=&S_0\phi(\overline{d}_1)-KP_0\phi(\overline{d}_2),\label{eq:52-1}\\
\overline{P}(K, T)&=&KP_0\phi(-\overline{d}_2)-S_0\phi(-\overline{d}_1).
\label{eq:53}
\end{eqnarray}
where
\begin{eqnarray}
P_0&=&\exp\Bigg\{-r_0T+\frac{2HT^{(\alpha-1)2H+2H+1}}{(\Gamma(\alpha))^{2H}((\alpha-1)2H+2H)((\alpha-1)2H+2H+1)}\nonumber\\
&\times&\left(\frac{\sigma_r^2T}{(\alpha-1)2H+2H+2}-\mu_r\right)\Bigg\}\\
\overline{d}_1&=&\frac{\ln\frac{S_0}{K}+\overline{r}T+\frac{1}{2}\overline{\sigma}^2T}{\overline{\sigma}\sqrt{T}},\\
\overline{d}_2&=&\overline{d}_1-\overline{\sigma}\sqrt{T},\\
\overline{r}&=&r_0+\frac{2HT^{(\alpha-1)2H+2H}}{(\Gamma(\alpha))^{2H}((\alpha-1)2H+2H)((\alpha-1)2H+2H+1)}\\
&\times&\left(\mu_r-\frac{\sigma_r^2T}{(\alpha-1)2H+2H+2}\right),\nonumber\\
\overline{\sigma}^2&=&\frac{2HT^{(\alpha-1)2H+2H-1}}{(\Gamma(\alpha))^{2H}((\alpha-1)2H+2H)}\Bigg(\sigma_s^2+\frac{\rho\sigma_r\sigma_sT}{(\alpha-1)2H+2H+1}\nonumber\\
&+&\frac{\sigma_r^2T^2}{((\alpha-1)2H+2H+1)((\alpha-1)2H+2H+2)}\Bigg).
\label{eq:54}
\end{eqnarray}
and $\phi(.)$ is the cumulative normal distribution function.
\end{cor}
Now, for an illustration of the differences among these models: the
Merton , subdiffusive Merton and our fractional Merton $(FM)$ and subdiffusive fractional Merton $(SFM)$ models, we report the theoretical prices of some
hypothetical options using different methods. The prices computed by different models are presented in
Table \ref{table:1}, where $S_0$ denotes the stock price, $P_M$ denotes
the prices computed by the Merton model, $P_{SM}$ denotes the price simulated
by the subdiffusive Merton model, $P_{FM}$ shows the price obtained by the $FM$ model and  $P_{SFM}$ denotes the price computed according to $SFM$ model.

\begin{table}[H]
\centering

\caption{Results by different pricing models. Here, $\alpha=0.9, H=0.6, K=3, \sigma_r=0.3, \sigma_s=0.4, \rho=0.4, \mu_r=0.5, r_0=0.3, T=0.3, t=0$.}
\begin{tabular}{|c c c c c c c c c|}
 \hline
&&$T=0.2$&&&&$T=1$ &&\\
\cline{2-4}\cline{6-8}

S   & $P_{M}$  & $P_{SM}$ & $P_{FM}$ &$P_{SFM}$ & $P_{M}$  & $P_{SM}$ & $P_{FM}$ &$P_{SFM}$ \\
\hline
2      & 0.0174&  0.0334&  0.0012&  0.0036     & 1.8826&  1.9129&  1.7986&  1.8347 \\
2.25   & 0.0638&  0.0979&  0.0122&  0.0236     & 2.1326&  2.1629&  2.0486&  2.0847 \\
2.5    &0.1598&   0.2126&  0.0587&  0.0859     & 2.3826&  2.4129&  2.2986&  2.3347 \\
2.75   &0.3094&   0.3754&  0.1687&  0.2094     & 2.6326&  2.6629&  2.5486&  2.5847 \\
3      &0.5023&   0.5752&  0.3440&  0.3900     & 2.8826&  2.1929&  2.7986&  2.8347  \\
3.25   &0.7235&   0.7988&  0.5630&  0.6086      & 3.1326&  3.1629&  3.0486&  3.0847 \\
3.5    &0.9604&   1.0360&  0.8026&  0.8466      & 3.3826&  3.4129&  3.2986&  3.3347 \\
3.75   &1.2094&   1.2801&  1.0498&  1.0926      & 3.6326&  3.6629&  3.5486&  3.5847 \\
4      &1.4527&   1.5275&  1.2991&  1.3414      & 3.8826&  3.9129&  3.7986&  3.8347 \\

\hline
\end{tabular}
\label{table:1}
\vspace{-2mm}

\end{table}

By comparing columns $P_M$, $P_{SM}$, $P_{FM}$ and $P_{SFM}$ in Table \ref{table:1}, we have the conclusion that the call option prices
obtained by four valuation models are close to each other in the both in-the-money and out-of-the-money cases with low and high maturities. Meanwhile, we
can see that the prices given by the our $FM$ and $SFM$ models are smaller than the
prices given by the Merton and subdiffusive Merton models \cite{cui2010comment, guo2017option}.

\begin{figure}[H]
  \centering
          \includegraphics[width=1\textwidth]{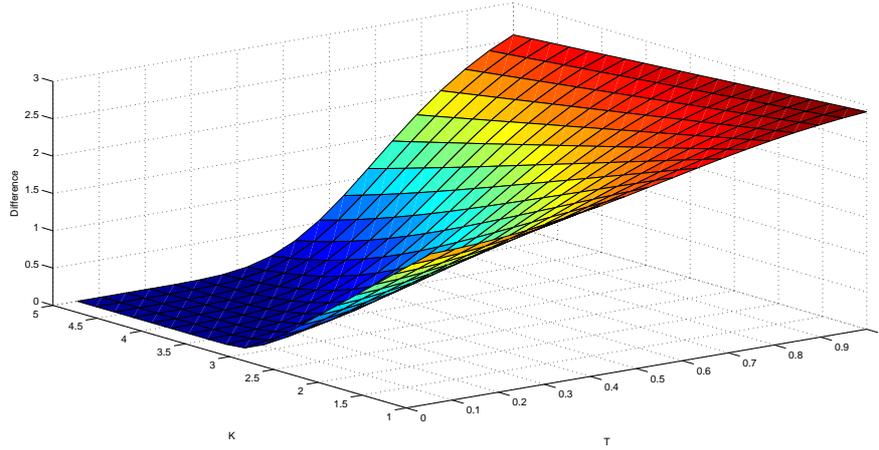}
  \caption{The European call option under $SFM$. Where  $r_0=0.1, \alpha=0.9, H=0.8, \sigma_r=0.3, \sigma_s=0.4, S_0=3, \mu_r=0.2, \rho=0.2$.}
\label{fig1}
\end{figure}

\begin{figure}[H]
  \centering
          \includegraphics[width=1\textwidth]{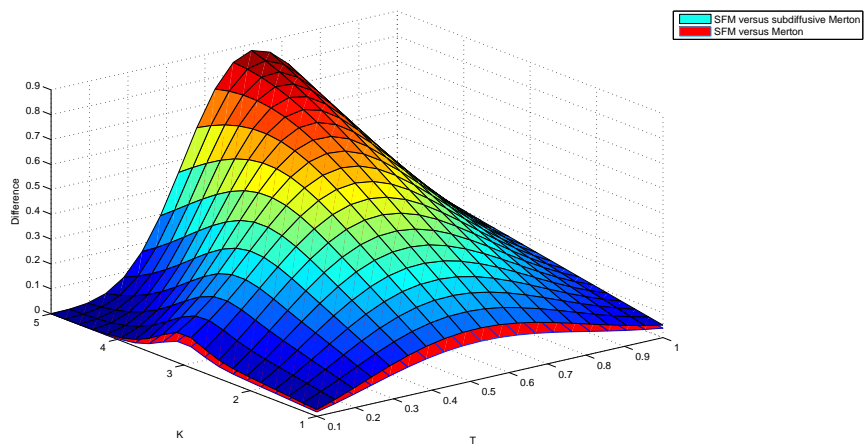}
  \caption{The difference between the price of the European call option under $SFM$, subdiffusive Merton and Merton models. Where  $r_0=0.1, \alpha=0.9, H=0.8, \sigma_r=0.3, \sigma_s=0.4, S_0=3, \mu_r=0.2, t=0, \rho=0.3$.}
\label{fig2}
\end{figure}

\begin{figure}[H]
  \centering
          \includegraphics[width=1\textwidth]{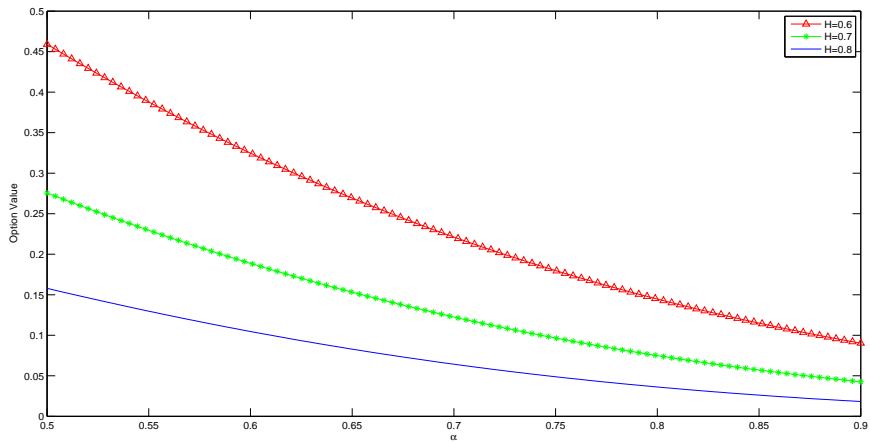}
  \caption{The European call option under $SFM$. Where  $r_0=0.3, \sigma_r=0.1, \sigma_s=0.3, S_0=4, \mu_r=0.2, \rho=0.2, t=0, T=0.2$.}
\label{fig3}
\end{figure}

We give three figures of the prices of European call options in the $SFM$ model for different parameters (see Figs. \ref{fig1}, \ref{fig2} and
\ref{fig3}). From Equations (\ref{eq:52-1})-(\ref{eq:54}), it is easy to see that $\sigma_{im}$ is just the implied volatility of the classical $BS$ model.
\section{Conclusion}\label{sec:5}

Previous option pricing research typically assumes that the risk-free rate or the short rate is constant during the life of the option. Since
fractional Brownian motion is a well-developed mathematical model
of strongly correlated stochastic processes, in this paper, we incorporate the fractional version of the Merton model with the subdiffusive mechanism to get better subdiffusive characteristic of financial markets. Then, we obtain pricing formula for call and put options when the short rate follows the subdiffusive fractional Merton model short rate and present some simulation results.

\bibliographystyle{siam}
\bibliography{../../reference}

\end{document}